\documentstyle[aps,eqsecnum]{revtex}
\begin{document}

\def\bdk{\mbox{\boldmath $k$}}
\def\bdx{\mbox{\boldmath $x$}}
\def\bdy{\mbox{\boldmath $y$}}
\def\bdyo{\mbox{\boldmath $y_1$}}
\def\bdyt{\mbox{\boldmath $y_2$}}
\def\bdpsi{\mbox{\boldmath $\psi$}}
\def\bdgdotpsi{\mbox{\boldmath $\gamma \cdot \psi$}}
\def\gdotpart{\mbox{\boldmath $\gamma \cdot \partial$}}
\def\bdkdotpsi{\mbox{\boldmath $k \cdot \psi$}}
\def\bdkdotg{\mbox{\boldmath $k \cdot \gamma$}}
\def\bdkdotx{\mbox{\boldmath $k \cdot x$}}
\def\bdkhdotpsi{\mbox{\boldmath $\hat{k} \cdot \psi$}}
\def\bdkhdotg{\mbox{\boldmath $\hat{k} \cdot \gamma$}}
\def\bdxydotg{\mbox{\boldmath ($x-y$)$\cdot \gamma$}}
\def\xdotg{\mbox{\boldmath $x \cdot \gamma$}}
\def\gdoty1y2{\mbox{\boldmath $\gamma \cdot$($y_1-y_2$)}}
\def\tpsi{\tilde{\psi}}
\def\tpsii{\tilde{\psi}_i}
\def\tpsij{\tilde{\psi}_j}
\def\tpsiz{\tilde{\psi}_0}
\def\tpsibar{\widetilde{\bar{\psi}}}
\def\tpsibarbold{\mbox{\boldmath $\widetilde{\bar{\psi}}$}}
\def\psibold{\mbox{\boldmath $\psi$}}
\def\psibarbold{\mbox{\boldmath $\bar{\psi}$}}
\def\cdotg{\mbox{\boldmath $\cdot \gamma$}}
\def\cdotG{\mbox{\boldmath $\cdot \Gamma$}}
\def\Gcdot{\mbox{\boldmath $\Gamma \cdot$}}
\def\gcdot{\mbox{\boldmath $\gamma \cdot$}}
\def\cdotk{\mbox{\boldmath $\cdot k$}}
\def\cdotkh{\mbox{\boldmath $\cdot \hat{k}$}}
\def\tpsiibar{\widetilde{\bar{\psi}}_i}
\def\tpsijbar{\widetilde{\bar{\psi}}_j}
\def\tpsizbar{\widetilde{\bar{\psi}}_0}
\def\bdbarpsi{\mbox{\boldmath $\tilde{\bar{\psi}}$}}
\def\khdottpsibarpmk{\mbox{\boldmath $\widetilde{\bar{\psi}}$}^+(\epsilon,
-\bdk) \mbox{\boldmath$\cdot \hat{k}$}}
\def\gdottpsi{\mbox{\boldmath $\gamma \cdot \tilde{\psi}$}}
\def\gdottpsibar{\mbox{\boldmath $\widetilde{\bar{\psi}} \cdot \gamma$}}
\def\gdottpsibarm{\mbox{\boldmath $\widetilde{\bar{\psi}}^- \cdot \gamma$}}
\def\kdottpsi{\mbox{\boldmath $k \cdot \tilde{\psi}$}}
\def\kdottpsibar{\mbox{\boldmath $\widetilde{\bar{\psi}} \cdot k$}}
\def\khdottpsi{\mbox{\boldmath $\hat{k} \cdot \tilde{\psi}$}}
\def\khdottpsibar{\mbox{\boldmath $\widetilde{\bar{\psi}} \cdot \hat{k}$}}
\def\khdottpsibarm{\mbox{\boldmath $\widetilde{\bar{\psi}}^- \cdot \hat{k}$}}
\def\khdottpsibarp{\mbox{\boldmath $\widetilde{\bar{\psi}}^+ \cdot \hat{k}$}}
\def\xdoty{\mbox{\boldmath $x \cdot y$}}
\def\bdpsib{\mbox{\boldmath $\bar{\psi}$}}
\def\bdxmy{\mbox{\boldmath ($x-y$)}}
\def\bdcdot{\mbox{\boldmath $\cdot$}}
\def\bdgamma{\mbox{\boldmath $\gamma$}}
\def\xidotpsi{\mbox{\boldmath $\bar{\xi}$}^-(\bdx)\mbox{\boldmath 
$\cdot \psi$}^-(\epsilon,\bdx)}
\def\psidotxi{\mbox{\boldmath $\bar{\psi}$}^+(\epsilon,\bdx)\mbox{\boldmath 
$\cdot \xi$}^+(\bdx)}
\def\psidotpsi{\mbox{\boldmath $\bar{\psi}$}(\epsilon,\bdx)\mbox{\boldmath 
$\cdot \psi$}(\epsilon,\bdx)}
\def\psidotpsimom{\mbox{\boldmath $\widetilde{\bar{\psi}}$}(\epsilon,-\bdk)
\mbox{\boldmath $\cdot \tilde{\psi}$}(\epsilon,\bdk)}
\def\psiggpsimom{\mbox{\boldmath $\widetilde{\bar{\psi}}$}(\epsilon,-\bdk)
\mbox{\boldmath $\cdot \gamma \, \gamma \cdot \tilde{\psi}$}(\epsilon,\bdk)}
\def\khat{\hat{k}}

\draft

\title{The massless gravitino and the $AdS$/CFT correspondence} 

\author{Steven Corley\thanks{scorley@phys.ualberta.ca}}

\address{Theoretical Physics Institute,
Department of Physics, University of Alberta,
Edmonton, Alberta, Canada T6G 2J1}

\maketitle
\begin{abstract}
We solve the Dirichlet boundary value problem for the massless gravitino
on $AdS_{d+1}$ space and compute the two-point function of the dual CFT
supersymmetry currents using the $AdS$/CFT correspondence principle.  We
find analogously to the spinor case that the boundary data for the
massless  $(d+1)$ dimensional bulk gravitino field consists of only a
$(d-1)$ dimensional gravitino.
\end{abstract}

\section{Introduction}

Recently Maldacena \cite{maldacena} has conjectured that the large N limit
of certain $d$ dimensional
conformal field theories is dual to supergravity or string
theory on $d+1$ dimensional Anti-de Sitter ($AdS$) space times
a compact manifold.  A prescription for generating correlators
of operators in the conformal field theory (CFT) from solutions
of the supergravity equations of motion has been given in 
\cite{gkp,wittone}.  The prescription associates to each field
$\phi_i$ in the supergravity action a corresponding local operator
${\cal O}^i$ in the CFT such that the following relation (in
Euclidean space) holds:
\begin{equation}
e^{-S_{eff}(\phi_i)} = \Bigl\langle e^{\int_{\partial} \phi_{i,0} {\cal O}^i}
\Bigr\rangle.
\label{correspondence}
\end{equation}
The effective action $S_{eff}$ is evaluated on the solutions to the
supergravity equations of motion subject to the boundary conditions
$\phi_i |_{\partial} = \phi_{i,0}$ where $\partial$ denotes the boundary
of $AdS_{d+1}$ space.  On the right-hand-side of (\ref{correspondence})
the expectation value of the given exponential is taken in the dual
conformal field theory, with $\phi_{i,0}$ acting as a source for
the CFT operator ${\cal O}^i$.  Using this relation various
two-point \cite{gkp}-\cite{Freedetal} and three and four-point 
\cite{Freedetal}-\cite{Freedetal2}
correlation functions have been computed, including detailed checks
of various Ward identities.

One subtlety of the prescription (\ref{correspondence}) involves the
manner in which $S_{eff}(\phi_i)$ is evaluated on the supergravity solutions.
Specifically $S_{eff}(\phi_i)$ diverges and must be regularized.  As
a result ambiguities in 
the overall coefficient of CFT correlators obtained from 
(\ref{correspondence}) arise, and therefore the CFT Ward identities
may not be satisfied.  However a regularization procedure which produces
correlators satisfying the Ward identities has been found in \cite{Freedetal}.
The procedure involves solving the Dirichlet boundary value problem for the
supergravity fields for a deformed boundary of $AdS$ such that
$S_{eff}$ is well-defined and only after obtaining the CFT correlators 
is the limit back to the true boundary of $AdS$ taken.

In this paper we consider the $AdS$/CFT correspondence for the massless
gravitino whose dual CFT operator is the supersymmetry
current \cite{FerrFrons}.  In section 2 
we solve the massless gravitino equations
of motion on the $AdS_{d+1}$ background
following the techniques of \cite{MuckVis2}.  We find, in
analogy to
the spinor case \cite{HS,MuckVis2}, that the boundary data
for the massless $(d+1)$ dimensional bulk gravitino field consists of only 
a $(d-1)$
dimensional gravitino due to the first order nature of the equation
of motion.  In section 3 we use the 
correspondence (\ref{correspondence}) to compute the two-point function
for the dual supersymmetry currents, taking care to evaluate the 
gravitino action in the manner discussed above,
and find the expected result.
In section 4 we make some concluding remarks.

\section{Constructing the solution}

In this section we solve the boundary value problem for the massless
Rarita-Schwinger field.  Our method for solving the equation of
motion parallels that of \cite{MuckVis2}.
We first find the most general solution to the Fourier transformed
equation of motion.
This solution contains an exponentially growing mode in $k$ and
therefore is not Fourier transformable.  
Demanding Fourier transformability we are forced to constrain the
boundary data such as to remove the undesired mode.  We then re-express
the solution in terms of the desired boundary data and Fourier
transform back to position space obtaining the bulk values of the
Rarita-Schwinger field.

The action for the massive Rarita-Schwinger field in $d+1$ dimensions is
given by
\begin{equation}
S = \frac{1}{2} \int d^{d+1}x \, e \, (\bar{\psi}_{\mu} \Gamma^{\mu \nu \rho}
D_{\nu} \psi_{\rho} + m_1 \bar{\psi}_{\mu} \psi^{\mu} + m_2 \bar{\psi}_{\mu}
\Gamma^{\mu \nu} \psi_{\nu})
\label{action}
\end{equation}
where
\begin{equation}
D_{\nu} \psi_{\rho} = (\partial_{\nu} + \frac{1}{4} 
\omega_{\nu}^{ab} \gamma_{ab})\psi_{\rho}
\end{equation}
and $m_1$ and $m_2$ are related to the mass $m$ and cosmological
constant $\Lambda$ (see \cite{ads5s5} and \cite{ads4s7} for the relation
for supergravity on $AdS_5 \times S_5$ and $AdS_7 \times S_4$ respectively).
Our notation is as follows: $e_{a}^{\mu}$ is the vielbein and $e$
its' determinant, $\omega_{\mu}^{ab}$ is the spin connection, $\Gamma^{\mu}$
are the curved space gamma matrices related to the flat space gamma matrices
$\gamma^{a}$ by $\Gamma^{\mu} = e_{a}^{\mu} \gamma^{a}$, the flat space
gamma matrices satisfy the anticommutation relations $\{\gamma^a,\gamma^b\}
= 2 \delta^{ab}$, gamma matrices
with more than one index are antisymmetrized as $\Gamma^{\mu_1 \cdots \mu_n}
= \Gamma^{[\mu_1} \cdots \Gamma^{\mu_n]}$, coordinate indices are denoted
by lower case Greek letters $\mu, \nu,...$
running from $0$ to $d$ and lower case Latin letters
$i,j,...$ running from $1$ to $d$, and lower case Latin letters $a, b,...$
denote Lorentz indices.
Varying the action (\ref{action}) with respect to $\bar{\psi}_{\mu}$
results in the equation of motion 
\begin{equation}
\Gamma^{\mu \nu \rho} D_{\nu} \psi_{\rho} + m_1 \psi^{\mu} + m_2
\Gamma^{\mu \nu} \psi_{\nu} =0
\label{eom}
\end{equation}
while varying with respect to $\psi_{\rho}$ results in the equation
of motion for the adjoint Rarita-Schwinger field
\begin{equation}
\bar{\psi}_{\mu} \loarrow{D}_{\nu} \Gamma^{\mu \nu \rho} - m_1
\bar{\psi}^{\rho} - m_2 \bar{\psi}_{\mu} \Gamma^{\mu \rho} = 0
\label{adjointeom}
\end{equation}
where $\loarrow{D}_{\nu} = \loarrow{\partial}_{\nu} - 
(1/4) \omega_{\nu}^{ab} \gamma_{ab}$.

We now specialize to the Euclidean $AdS_{d+1}$ background geometry and choose
coordinates such that the metric takes the form
\begin{equation}
ds^2 = \frac{1}{(x^0)^2} \bigl((dx^0)^2 + d{\bf x} \cdot d{\bf x} \bigr)
\label{metric}
\end{equation}
where we use boldface letters to denote the inner
product of 
$d$-dimensional vectors in the flat Euclidean metric, 
$\xdoty:= x_i \delta^{ij} y_j$, and use the flat Euclidean metric
to raise and lower indices, $x^i = \delta^{ij}x_j$.
Choosing the corresponding vielbein to be 
\begin{equation}
e^{a}_{\mu} = 
\frac{1}{x^0} \delta^{a}_{\mu}
\label{vielbein}
\end{equation}
it is straightforward to show that the spin 
connection is given by
\begin{equation}
\omega_{\mu}^{ab} = -\frac{1}{x^0} (\delta^{a}_{\mu} \delta^{b}_{0}
-\delta^{b}_{\mu} \delta^{a}_{0}),
\label{spinconn}
\end{equation}
as is easily verified by substituting into $de^a + \omega^{ab} \wedge
e_b=0$.

To solve the equation of motion (\ref{eom}) it is first convenient
to rewrite it in the form
\begin{equation}
\Gamma^{\nu} (D_{\nu} \psi_{\rho} - D_{\rho} \psi_{\nu}) + m_- \psi_{\rho}
- \frac{m_+}{d-1} \Gamma_{\rho} \Gamma^{\nu}\psi_{\nu}=0
\label{eom2}
\end{equation}
where $m_{\pm} = m_1 \pm m_2$.
Expanding (\ref{eom2}) and substituting (\ref{vielbein}) for the vielbein 
and (\ref{spinconn}) for the spin connection we find the $\rho=0$ equation
\begin{equation}
\biggl( x^0 \partial_0 + \frac{m_+}{d-1} \gamma_0 \biggr) \bdgdotpsi =
\biggl( x^0 \gdotpart - \frac{d}{2} \gamma_0 + m_- - \frac{m_+}{d-1}\biggr)
\psi_0
\label{zero}
\end{equation}
and the $\rho=i$ equation
\begin{eqnarray}
\biggl(x^0 (\gamma^0 \partial_0 + \gdotpart) - \Bigl(\frac{d}{2} -1 \Bigr)
\gamma_0 + m_- \biggr) \psi_i = \biggl(x^0 \gamma^0 \partial_i + \frac{1}{2}
\gamma_i 
+ \frac{m_+}{d-1} \gamma_i \gamma^0 \biggr) \psi_0
+ \biggl( x^0 \partial_i - \frac{1}{2} \gamma_i \gamma_0 
+ \frac{m_+}{d-1} \gamma_i \biggr) \bdgdotpsi
\label{one}
\end{eqnarray}
To solve these equations we work in momentum space. 
Define the Fourier transform
\begin{equation}
\psi_{\mu}(x^0,\bdx) = \frac{1}{(2 \pi)^{d/2}} \int d^dk \,
e^{i \bdkdotx} \tilde{\psi}_{\mu}(x^0,\bdk)
\end{equation}
and substitute into (\ref{zero}) and (\ref{one}).  We obtain after
some rearranging
\begin{equation}
\biggl(x^0\partial_0 + \frac{m_+}{d-1} \gamma_0 \biggr) \gdottpsi 
= \biggl(i x^0 \bdkdotg - \frac{d}{2} \gamma^0 + 
m_- - \frac{m_+}{d-1} \biggr) \tpsiz
\label{zero'} 
\end{equation}
\begin{equation}
\biggl(x^0 \partial_0 - i x^0 \bdkdotg \gamma^0 - \Bigl(\frac{d}{2} -1 \Bigr) 
+ m_- \gamma^0 \biggr) \tpsii = \biggl(ix^0 k_i \gamma^0 +\frac{1}{2} \gamma_i 
-\frac{m_+}{d-1} \gamma_i \gamma^0 \biggr)
\gdottpsi + \biggl(i x^0 k_i -\frac{1}{2} \gamma_i \gamma^0
- \frac{m_+}{d-1} \gamma_i \biggr)\tpsiz
\label{one'}
\end{equation}
for the $\rho=0$ and $\rho=i$ equations respectively.

Solving equations (\ref{zero'},\ref{one'}) is straightforward but tedious.
The resulting solution for the generic case presented so far is complicated
and therefore we consider the simpler massless case $m=0$ in the remainder.
This is equivalent to demanding $m_+ = - m_- = \Lambda/2$ ($\Lambda$ here
is related to the cosmological constant by a dimension dependent factor).  
To simplify the equations above we start by defining the projection operator
\begin{equation}
P_{i}^j := \delta_{i}^j - \frac{\khat_i}{d-1} \biggl(d \khat^j - \bdkhdotg
\gamma^j \biggr) - \frac{\gamma_i}{d-1} \biggl(\gamma^j - \bdkhdotg 
\khat^j \biggr)
\label{proj1}
\end{equation}
which is orthogonal to both $k^i$ and $\gamma^i$.  
Defining the tranverse components of $\tpsii$ to $k^i$ and $\gamma^i$ 
respectively as $\tpsii^T := P_{i}^j \tpsij$ it follows that the
field may be decomposed as
\begin{equation}
\tpsii = \tpsii^T + \frac{1}{d-1}(\gamma_i - \khat_i \bdkhdotg) \gdottpsi
+\frac{1}{d-1}(d \khat_i - \gamma_i \bdkhdotg) \khdottpsi.
\label{decomp}
\end{equation}
The equation of motion for $\tpsii^T$ is easily derived by applying
$P_{i}^j$
to (\ref{one'}) obtaining
\begin{equation}
\biggl(x^0 \partial_0 - i x^0 \bdkdotg \gamma^0 - \Bigl(\frac{d}{2} -1 \Bigr) 
- \frac{\Lambda}{2} \gamma^0 \biggr) \tpsii^T = 0.
\label{tpsiiTeom}
\end{equation}

Finding the equation of motion for $\gdottpsi$ is a little more involved.
First solve (\ref{zero'}) for $\tpsiz$ and substitute into
(\ref{one'}) contracted with $\gamma^i$.  This results in the
algebraic relation\footnote{Here is the essential between the massless and
massive cases.  In the massive case $\kdottpsi$ cannot be expressed
algebraically in terms of $\gdottpsi$.}
\begin{equation}
\kdottpsi = \biggl(\bdkdotg + i \frac{(d-1)\gamma^0 - \Lambda}{2 x^0}
\biggr) \gdottpsi.
\label{kdottpsi}
\end{equation}
Substituting this relation for $\kdottpsi$ into (\ref{one'}) contracted
with $k^i$ and comparing with (\ref{zero'}) results in the
second algebraic relation
\begin{equation}
\tpsiz = - \gamma^0 \gdottpsi.
\label{tpsiz}
\end{equation}
From this and (\ref{zero'}) the equation of motion for $\gdottpsi$
follows
\begin{equation}
\biggl(x^0 \partial_0 + i x^0 \bdkdotg \gamma^0 -\frac{d}{2}
- \frac{\Lambda}{2} \gamma^0 \biggr) \gdottpsi = 0
\label{gdottpsieom}
\end{equation}

The equations of motion (\ref{tpsiiTeom},\ref{gdottpsieom}) are
easily integrated in the form of path ordered exponentials.  The 
exponentials are straightforward to evaluate and we find
\begin{eqnarray}
\tpsii^T(x^0,\bdk) 
& = & (x^0)^{(d-1)/2} \biggl( (1+\gamma^0) \Bigl(I_{(\Lambda-1)/2}(k x^0)
+ i \bdkdotg \gamma^0 I_{-(\Lambda-1)/2}(k x^0) \Bigr) \nonumber \\
& + & (1-\gamma^0) \Bigl(I_{-(\Lambda+1)/2}(k x^0) +
i \bdkdotg \gamma^0 I_{(\Lambda+1)/2}(k x^0) \Bigr) \biggr) \lambda_i(\bdk)
\label{tpsiisol}
\end{eqnarray}
where $\lambda_i(\bdk)$ is transverse to both $k^i$ and $\gamma^i$ and
\begin{eqnarray}
\gdottpsi(x^0,\bdk) 
& = & (x^0)^{(d+1)/2} \biggl( (1-\gamma^0) \Bigl(I_{(\Lambda-1)/2}(k x^0)
- i \bdkdotg \gamma^0 I_{-(\Lambda-1)/2}(k x^0) \Bigr) \nonumber \\
& + & (1+\gamma^0) \Bigl(I_{-(\Lambda+1)/2}(k x^0) -
i \bdkdotg \gamma^0 I_{(\Lambda+1)/2}(k x^0) \Bigr) \biggr) \lambda_0(\bdk).
\label{gdottpsisol}
\end{eqnarray}
Using the algebraic relation expressing $\khdottpsi$ in terms
of $\gdottpsi$ we now have the complete momentum space solution for
the massless Rarita-Schwinger field.  This solution however is not
Fourier transformable since the modified Bessel function $I_{\nu}(k x^0)$
diverges exponentially for large $k$.  In order to obtain a Fourier
transformable solution we must constrain the boundary fields
$\lambda_i(\bdk)$ and $\lambda_0(\bdk)$ to remove the growing
modes.  The same problem occurs for the spinor field as discussed
in \cite{MuckVis2} and more generally for any field satisfying
a first order equation of motion.  The simple reason is that
for a first order equation only the boundary values of the field
are freely specifiable, whereas for a second order equation of motion
the boundary value and first derivative of the field are freely
speciable or conversely the boundary value and asymptotic behavior.
As emphasized in \cite{MuckVis2} the asymptotic behavior of the
field is essential for the $AdS$/CFT correspondence and while
specifying it for a second order equation of motion is not a
problem, for a first order equation it must be done at the expense
of specifying completely the boundary values of the field.

We may easily find the constraints necessary to remove the 
offending terms by rewriting the solutions (\ref{tpsiisol},
\ref{gdottpsisol}) in terms of the modified Bessel
functions $I_{(\Lambda \pm 1)/2}(k x^0)$ and $K_{(\Lambda \pm 1)/2}(k x^0)$
where $K_{(\Lambda \pm 1)/2}(k x^0)$ decay exponentially for large $k$.
The conditions necessary for removing the exponentially growing modes are
easily shown to be
$(1 + i \bdkhdotg \gamma^0) \lambda_i(\bdk) = 0$
and
$(1 - i \bdkhdotg \gamma^0) \lambda_0(\bdk) = 0$
which effectively removes half of the components of $\lambda_i$ and
$\lambda_0$ respectively.  Defining $\lambda_{i}^{\pm} := 
(1/2)(1 \pm \gamma^0) \lambda_i$ and similarly for
$\lambda_0$ these conditions may be rewritten in the equivalent form
\begin{eqnarray}
\lambda_{i}^+ = i \bdkhdotg \lambda_{i}^- \\
\lambda_{0}^+ = - i \bdkhdotg \lambda_{0}^-.
\end{eqnarray}
Substituting these relations into the solutions (\ref{tpsiisol},
\ref{gdottpsisol}) results in\footnote{These solutions may equivalently
be expressed in terms of $\lambda_{i}^+$ and $\lambda_{0}^+$.}
\begin{eqnarray}
\tpsii^T(x^0,\bdk) = (x^0)^{(d-1)/2} \Bigl( K_{(\Lambda+1)/2}(k x^0)
+ i \bdkhdotg K_{(\Lambda-1)/2}(k x^0) \Bigr) \lambda_{i}^-(\bdk)
\label{tpsiir} \\
\gdottpsi(x^0,\bdk) = (x^0)^{(d+1)/2} \Bigl( K_{(\Lambda+1)/2}(k x^0)
- i \bdkhdotg K_{(\Lambda-1)/2}(k x^0) \Bigr) \lambda_{0}^-(\bdk)
\label{gdottpsir}
\end{eqnarray}
where constants have been absorbed into $\lambda_i$ and $\lambda_0$
respectively.

The final step before inverse Fourier transforming back to position space
is to re-express the solutions (\ref{tpsiir},\ref{gdottpsir}) in terms
of the given boundary data $\tpsii(\epsilon,\bdk)$.   Clearly the 
components of the boundary Rarita-Schwinger field are not all
independent as discussed above.
To find the independent components evaluate (\ref{tpsiir}) at the
boundary $x^0 = \epsilon$
(we consider
only the $\tpsii^T$ case in detail here as the $\gdottpsi$ case is similar)
and apply the projection operators $(1/2)(1 \pm \gamma^0)$ deriving
the relations
\begin{eqnarray}
\tpsii^{T,-}(\epsilon,\bdk) & = & \epsilon^{(d-1)/2} K_{(\Lambda+1)/2}
(k \epsilon)
\lambda_{i}^-(\bdk) \label{lambdai} \\
\tpsii^{T,+}(\epsilon,\bdk) & = & \epsilon^{(d-1)/2}
i \bdkhdotg K_{(\Lambda-1)/2}(k \epsilon)
\lambda_{i}^-(\bdk). 
\end{eqnarray}
Solving one relation for $\lambda_i$ and substituting into the other yields
\begin{equation}
\tpsii^{T,+}(\epsilon,\bdk) = \frac{K_{(\Lambda-1)/2}(k \epsilon)}
{K_{(\Lambda+1)/2}(k \epsilon)} i \bdkhdotg \,
\tpsii^{T,-}(\epsilon,\bdk).
\end{equation} 
Using the small $z$ expansion of $K_{\nu}(z)$
\begin{equation}
K_{\nu}(z) = \frac{1}{2} \biggl(\Gamma(\nu) \Bigl( \frac{z}{2} \Bigr)^{-\nu}
\bigl(1 + {\cal O}(z^2)\bigr) + 
\Gamma(-\nu) \Bigl( \frac{z}{2} \Bigr)^{\nu}
\bigl(1 + {\cal O}(z^2)\bigr) \biggr)
\label{smalleps}
\end{equation}
we find that for regular boundary data in the $\epsilon \rightarrow 0$
limit we must demand that $\tpsii^{T,+}(\epsilon,\bdk) = 0$ and therefore
that the appropriate boundary data is given by 
$\tpsii^{T,-}(\epsilon,\bdk)$.  Eliminating $\lambda_i(\bdk)$
for $\tpsii^{T,-}(\epsilon,\bdk)$ in (\ref{lambdai}) 
yields the momentum space solution
for $\tpsii(x^0,\bdk)$
\begin{equation}
\tpsii^T(x^0,\bdk) = \Bigl( \frac{x^0}{\epsilon} \Bigr)^{(d-1)/2}
\frac{K_{(\Lambda+1)/2}(k x^0) + i \bdkhdotg K_{(\Lambda-1)/2}(k x^0)}
{K_{(\Lambda+1)/2}(k \epsilon)} \tpsii^{T,-}(\epsilon,\bdk)
\label{finaltpsiT}
\end{equation}
expressed in terms of the desired boundary data.
Similar manipulations with the $\gdottpsi$ solution (\ref{gdottpsir})
yields the relation between the chiral components of 
$\gdottpsi(\epsilon,\bdk)$
\begin{equation}
\gdottpsi^{-}(\epsilon,\bdk) = - i \bdkhdotg 
\frac{K_{(\Lambda-1)/2}(k \epsilon)}
{K_{(\Lambda+1)/2}(k \epsilon)} 
\gdottpsi^{+}(\epsilon,\bdk)
\label{gdottpsimp}
\end{equation}
and the resulting solution expressed in terms of the boundary data
$\gdottpsi^{+}(\epsilon,\bdk)$
\begin{equation}
\gdottpsi(x^0,\bdk) = \Bigl( \frac{x^0}{\epsilon} \Bigr)^{(d+1)/2}
\frac{K_{(\Lambda+1)/2}(k x^0) - i \bdkhdotg K_{(\Lambda-1)/2}(k x^0)}
{K_{(\Lambda+1)/2}(k \epsilon)} \gdottpsi^{+}(\epsilon,\bdk).
\label{gdottpsiinter}
\end{equation}

This is not quite the desired form for $\gdottpsi(x^0,\bdk)$ though.  
Recall that 
$\gdottpsi$ and $\kdottpsi$ are related as in (\ref{kdottpsi}).  Evaluating
this relation at $x^0 = \epsilon$ and using (\ref{gdottpsimp}) to solve
for $\gdottpsi^+$ yields
\begin{equation}
\gdottpsi^+(\epsilon,\bdk) = 2 \epsilon \frac{1 + i \bdkhdotg 
\bigl(K_{(\Lambda-1)/2}(k \epsilon)/K_{(\Lambda+1)/2}(k \epsilon)\bigr)}
{1+\bigl(K_{(\Lambda-1)/2}(k \epsilon)/K_{(\Lambda+1)/2}(k \epsilon)\bigr)^2}
\frac{2 \epsilon \bdkdotg + i \bigl( (d-1)\gamma^0 + \Lambda \bigr)}
{(2 \epsilon k)^2 -(d-1)^2 + \Lambda^2} \kdottpsi(\epsilon,\bdk)
\label{gdottpsiinkdottpsi}
\end{equation}
From the expansion (\ref{smalleps}) it follows that for regular
boundary data 
in the $\epsilon \rightarrow 0$ limit we must demand that
$\gdottpsi^+(\epsilon,\bdk) = 0$, and so is not the correct boundary
data.  Applying $(1/2)(1 + \gamma^0)$ to (\ref{gdottpsiinkdottpsi}) 
(which annihilates
the left-hand-side) one arrives at
\begin{equation}
\bdkhdotg \biggl(2 k \epsilon  - \Bigl( (d-1) \gamma^0 + \Lambda \Bigr)
\frac{K_{(\Lambda-1)/2}(k \epsilon)}{K_{(\Lambda+1)/2}(k \epsilon))}
\biggr) \khdottpsi^-(\epsilon,\bdk) = -i \biggl( \Bigl((d-1) \gamma^0 
+ \Lambda \Bigr) + 2 k \epsilon
\frac{K_{(\Lambda-1)/2}(k \epsilon)}{K_{(\Lambda+1)/2}(k \epsilon))}
\biggr) \khdottpsi^+(\epsilon,\bdk).
\label{kh+inkh-}
\end{equation}
Using the expansion (\ref{smalleps}) once again we see that the correct
boundary data is $\khdottpsi^-(\epsilon,\bdk)$.  We may now
express $\gdottpsi(x^0,\bdk)$ in terms of this boundary data by
substituting (\ref{gdottpsiinkdottpsi}) in (\ref{gdottpsiinter}) 
with $\khdottpsi(\epsilon,\bdk)$ given in terms of 
$\khdottpsi^-(\epsilon,\bdk)$ after using (\ref{kh+inkh-}).
Because the result is quite complicated we do not give the expression
explicitly.

Finally we may inverse Fourier transform to find the bulk value
of the field in position space.  This is not as formidable
as it might first appear because we need only work to leading
order in $\epsilon$.  Substituting the solutions for $\tpsii^T(x^0,\bdk)$
(\ref{finaltpsiT}) and $\gdottpsi(x^0,\bdk)$ (\ref{gdottpsiinter},
\ref{gdottpsiinkdottpsi},\ref{kh+inkh-}) into the decomposition (\ref{decomp}),
expanding the $\epsilon$ dependent
terms and simplifying results in 
\begin{eqnarray}
\tpsii(x^0,\bdk) & = & \frac{2^{-(\Lambda-1)/2}}
{\Gamma \bigl((1+\Lambda)/2 \bigr)}
\epsilon^{(\Lambda - d)/2+1} (x^0)^{(d-1)/2} k^{(\Lambda+1)/2}
\biggl( 
\bigl(K_{(\Lambda+1)/2}(k x^0) + i \bdkhdotg K_{(\Lambda-1)/2}(k x^0) \bigr) 
\delta_{i}^j \nonumber \\
& + & 
 \frac{2}{d-1+\Lambda} \Bigl(
\bigl(K_{(\Lambda-1)/2}(k x^0) + i \bdkhdotg K_{(\Lambda+1)/2}(k x^0) 
\bigr) k x^0
\khat_i \khat^j \nonumber \\
& - & 
K_{(\Lambda-1)/2}(k x^0) \bigl(i \gamma_i + 
(\Lambda-1) i \bdkhdotg \khat_i \bigr)\khat^j
\Bigr) \biggr) \tpsij^-(\epsilon,\bdk).
\end{eqnarray}
The Fourier transform of this expression is tedious to carry out, but
straightforward, and we arrive at
\begin{equation}
\psi_i(x^0,\bdx) = - c (x^0)^{(d+\Lambda)/2-1}
\int d^dy \, \frac{x^0 \gamma^0 + \bdxydotg}{\bigl( (x^0)^2 + |\bdx - \bdy|^2
\bigr)^{(d+\Lambda+1)/2}} \biggl(\delta_{i}^j - 2 \frac{(x-y)_i
(x-y)^j}{(x^0)^2 + |\bdx-\bdy|^2} \biggr) \psi^{-}_j(\epsilon,\bdy)
\label{xsoli}
\end{equation}
where
\begin{equation}
c = \frac{1}{\pi^{d/2}}
\frac{\Gamma\bigl( (d+1+\Lambda)/2 \bigr)}
{\Gamma\bigl( (1+\Lambda)/2 \bigr)}
\frac{d+1+\Lambda}
{d-1+\Lambda},
\label{ccoeff}
\end{equation}
a factor of $\epsilon^{(\Lambda-d)/2 +1}$ has been absorbed into
$\psi^{-}_j(\epsilon,\bdk)$, and 
$\psi^{-}_j(\epsilon,\bdk)$ is transverse to $\gamma^j$ as
follows from (\ref{gdottpsimp}) in the $\epsilon \rightarrow 0$ limit.  The
remaining field component is given by 
\begin{equation}
\psi_0(x^0,\bdx) =
- \gamma^0 \bdgdotpsi(x^0,\bdx).
\label{xsol0}
\end{equation}

For the adjoint Rarita-Schwinger field the analysis of solving
the equation of motion (\ref{adjointeom}) exactly parallels the
above discussion, therefore we only
quote the results.  The momentum space adjoint field may be
decomposed as
\begin{equation}
\tpsiibar = \tpsiibar^T + \khdottpsibar \frac{d \khat_i - \bdkhdotg \gamma_i}
{d-1} + \gdottpsibar \frac{\gamma_i - \bdkhdotg \khat_i}{d-1}
\label{decompadjoint}
\end{equation}
where $\tpsiibar^T := \tpsijbar P^{j}_{\, \, i}$ and
\begin{equation}
\kdottpsibar = \gdottpsibar \biggl(\bdkdotg + i \frac{(d-1)\gamma^0 + \Lambda}
{2 x^0} \biggr).
\end{equation}
The momentum space equations of motion for the the fields $\tpsiibar^T$
and $\gdottpsibar$ may be derived and solved in strict analogy to the
unbarred case and we find for the Fourier transformable part 
\begin{equation}
\tpsiibar^T(x^0,\bdk) = \tpsiibar^{T,+}(\epsilon,\bdk) 
\frac{K_{(\Lambda+1)/2}(k x^0) - i \bdkhdotg K_{(\Lambda-1)/2}(k x^0)}
{K_{(\Lambda+1)/2}(k \epsilon)}
\Bigl( \frac{x^0}{\epsilon} \Bigr)^{(d-1)/2}
\label{finaltpsibarT}
\end{equation}
and
\begin{equation}
\tpsibarbold(x^0,\bdk)\cdotg = \tpsibarbold^-(\epsilon,\bdk)\cdotg
\frac{K_{(\Lambda+1)/2}(k x^0) + i \bdkhdotg K_{(\Lambda-1)/2}(k x^0)}
{K_{(\Lambda+1)/2}(k \epsilon)}
\Bigl( \frac{x^0}{\epsilon} \Bigr)^{(d+1)/2}.
\label{gdottpsibarinter}
\end{equation}
$\tpsibarbold^-(\epsilon,\bdk)\cdotg$ is not the desired boundary data
to express $\tpsibarbold(x^0,\bdk)\cdotg$ in 
terms of as it goes to zero in the
limit $\epsilon \rightarrow 0$.  Rather we must express 
$\tpsibarbold(x^0,\bdk)\cdotg$
in terms of $\tpsibarbold^+(\epsilon,\bdk)\cdotkh$ which can be done after
using the relations
\begin{equation}
\tpsibarbold^-(\epsilon,\bdk)\cdotg
= \tpsibarbold(\epsilon,\bdk)\cdotk \,
2 \epsilon \,
\frac{2 \epsilon \bdkdotg + i \bigl( (d-1)\gamma^0 - \Lambda \bigr)}
{(2 \epsilon k)^2 -(d-1)^2 + \Lambda^2} 
\frac{1 - i \bdkhdotg 
\bigl(K_{(\Lambda-1)/2}(k \epsilon)/K_{(\Lambda+1)/2}(k \epsilon)\bigr)}
{1+ \bigl(K_{(\Lambda-1)/2}(k \epsilon)/K_{(\Lambda+1)/2}(k \epsilon)\bigr)^2}
\label{tpsibar-intpsibark}
\end{equation}
and
\begin{equation}
\tpsibarbold^+(\epsilon,\bdk)\cdotkh \, \bdkhdotg \biggl(2 k \epsilon  - 
\Bigl( (d-1) \gamma^0 + \Lambda \Bigr)
\frac{K_{(\Lambda-1)/2}(k \epsilon)}{K_{(\Lambda+1)/2}(k \epsilon))}
\biggr) = -i 
\tpsibarbold^-(\epsilon,\bdk)\cdotkh \biggl( \Bigl((d-1) \gamma^0 
- \Lambda \Bigr) - 2 k \epsilon
\frac{K_{(\Lambda-1)/2}(k \epsilon)}{K_{(\Lambda+1)/2}(k \epsilon))}
\biggr).
\label{tpsibar+intpsibar-}
\end{equation}
Substituting these expressions into (\ref{gdottpsibarinter}), keeping only
the leading order terms in $\epsilon$, and inverse Fourier
transforming yields
\begin{equation}
\bar{\psi}_i(x^0,\bdx) = c (x^0)^{(d+\Lambda)/2-1}
\int d^dy \, \bar{\psi}^{+}_j(\epsilon,\bdy) 
\frac{x^0 \gamma^0 + \bdxydotg}{\bigl( (x^0)^2 + |\bdx - \bdy|^2
\bigr)^{(d+\Lambda+1)/2}} \biggl(\delta_{i}^j - 2 \frac{(x-y)_i
(x-y)^j}{(x^0)^2 + |\bdx-\bdy|^2} \biggr)
\label{xbarsoli}
\end{equation}
where a factor of $\epsilon^{(\Lambda-d)/2+1}$ has been absorbed
into $\bar{\psi}^{+}_j(\epsilon,\bdy)$ and 
$\bar{\psi}^{+}_j(\epsilon,\bdy)$ is transverse to $\gamma^j$.
The remaining component of the adjoint field is given by 
\begin{equation}
\tpsizbar(x^0,\bdx) = - \tpsibarbold(x^0,\bdx)\cdotg \gamma^0.
\label{xbarsol0}
\end{equation}

\section{Two-point function}

The $AdS$/CFT correspondence is prescribed by the relation
\begin{equation}
\exp(-S) =\Bigl\langle \exp \bigl(\int_{\partial}
\xidotpsi + \psidotxi \bigr) \Bigr\rangle
\label{adscft}
\end{equation}
where $S$ is the action (\ref{action}) evaluated on the
solutions found in the previous section
with prescribed boundary data $\psi^{-}_i(\epsilon,\bdx)$ and 
$\bar{\psi}^{+}_i(\epsilon,\bdx)$.  The right-hand-side of (\ref{adscft})
is the expectation value of the given exponential in the
boundary conformal field theory with $\psi^{-}_i(\epsilon,\bdx)$ and
$\bar{\psi}^{+}_i(\epsilon,\bdx)$ playing the role of source terms for the
boundary conformal fields $\bar{\xi}^{-}_i(\bdx)$ and $\xi^{+}_i(\bdx)$ 
respectively.  We expect these boundary conformal fields to be the
supersymmetry currents of the boundary CFT \cite{FerrFrons} and 
indeed the two-point
function we find below is in agreement with this expectation.
The two-point function
$\langle \xi^{+}_i(\bdx) \bar{\xi}^{-}_j(\bdy) \rangle$ is 
easily obtained from (\ref{adscft}) by taking a pair of functional
derivatives with respect to $\psi^{-}_j(\epsilon,\bdy)$ 
and $\bar{\psi}^{+}_i(\epsilon,\bdx)$.
In our case the action (\ref{action}) is first order in derivatives
and therefore vanishes when evaluated on the solutions 
(\ref{xsoli},\ref{xsol0},\ref{xbarsoli},\ref{xbarsol0}).  This is also the
case for spinors as first noted by \cite{HS}.  To avoid this problem
\cite{HS} added the most general Lorentz invariant and generally covariant
boundary term quadratic in the spinor fields to the 
action\footnote{Justification for
this term has recently been given in \cite{arufrol} from a Hamiltonian
formulation and presumably extends to the gravitino case as well.}.  
Following this prescription we add to the action
principle (\ref{action}) the boundary term
\begin{equation}
S' =  \int_{\partial} d^dx \sqrt{h} \, \Bigl( a \bar{\psi}_i (\epsilon,\bdx) 
h^{ij}
\psi_j(\epsilon,\bdx) + b \psibarbold(\epsilon,\bdx) \cdotG \, \Gcdot
\psibold(\epsilon,\bdx) \Bigr)
\label{boundaryaction}
\end{equation}
where $h_{ij}$ is the induced metric on the boundary,
$h$ its' determinant and 
$a$ and $b$ are undetermined coefficients.  

To evaluate $S'$ the solutions (\ref{xsoli},\ref{xbarsoli}) could
be substituted and the integral performed.  However it is somewhat
simpler to work in momentum space where the boundary action becomes
\begin{equation}
S' = \epsilon^{-(d-2)} \int d^dk \,\Bigl(a \psidotpsimom + b \psiggpsimom 
\Bigr)
\label{bactionmom}
\end{equation}
where as before the dot product is defined using the Kronecker delta.
Substituting the decompositions (\ref{decomp}) and (\ref{decompadjoint}) for 
$\tpsii$ and $\tpsiibar$ respectively in the boundary action
(\ref{bactionmom}) we rewrite it as $S' = S_1 + S_2$
where
\begin{eqnarray}
S_1 & = & \frac{a}{\epsilon^{(d-2)}} \int d^dk \, \tpsiibar^T(\epsilon,-\bdk)
\delta^{ij} \tpsij^T(\epsilon,\bdk)
\label{S1} \\
S_2 & = & \frac{a}{\epsilon^{(d-2)}} \int d^dk \, 
\tpsibarbold(\epsilon,-\bdk)\cdotg
\biggl(1+ \frac{b}{a} + i \bdkhdotg \frac{(d-1)\gamma^0 - \Lambda}{k \epsilon}
+\frac{d}{d-1} \frac{(d-1)^2 - \Lambda^2}{(2 k \epsilon)^2} \biggr)
\gdottpsi(\epsilon,\bdk). \label{S2}
\end{eqnarray}
Substituting the solutions (\ref{finaltpsiT},
\ref{finaltpsibarT}) evaluated at $x^0 = \epsilon$ into $S_1$ and expanding
the projection operator we find
\begin{equation}
S_1 =2 \frac{a}{\epsilon^{(d-2)}} \int d^dk \, \tpsiibar^+(\epsilon,-\bdk)
\, i \bdkhdotg \frac{K_{(\Lambda-1)/2}(k \epsilon)}
{K_{(\Lambda+1)/2}(k \epsilon)}
\biggl(\delta^{ij} - \frac{d-2}{d-1} 
\khat^i \khat^j \biggr)\tpsij(\epsilon,\bdk)
\label{S1inter}
\end{equation}
In the limit of small $\epsilon$ the ratio of modified Bessel
functions may be expanded using (\ref{smalleps}) as
\begin{equation}
\frac{K_{(\Lambda-1)/2}(k \epsilon)}
{K_{(\Lambda+1)/2}(k \epsilon)} = \frac{k \epsilon}{\Lambda-1}
\Bigl(1 + \sum_{i=1}^{\infty} c_i k^{2i} \Bigr) +
\frac{\Gamma\bigl((1-\Lambda)/2\bigr)}
{\Gamma\bigl((1+\Lambda)/2\bigr)} \biggl( \frac{k \epsilon}
{2} \biggr)^{\Lambda} + \cdots
\label{ratioexp}
\end{equation}
where the $c_i$ are known coefficients which are not needed here.
Substituting this into (\ref{S1inter}) we see that from the first
term in the expansion
only the leading order
$k \epsilon$ term (multiplying $\khat^i \khat^j$)
gives rise to a non-trivial contribution to
$S_1$ while the remaining terms in parentheses give rise
to contact terms.  As we will see shortly though this term cancels
against a contribution from $S_2$, therefore the leading order
non-trivial contribution to $S_1$ actually comes from the
$(k \epsilon)^{\Lambda}$ term in the expansion\footnote{We are assuming
here that $\Lambda$ is not equal to an odd integer and
is $> 1$.  If $\Lambda$ is an odd integer 
then the expansion contains log terms as well
which give rise to the leading order non-trivial contributions to $S_1$
and $S_2$.
Nevertheless the two-point function obtained below is valid for
any $\Lambda > 0$.}.

Using (\ref{gdottpsiinter},\ref{gdottpsiinkdottpsi},\ref{kh+inkh-}) 
to express $\gdottpsi(\epsilon,\bdk)$ in terms of 
$\khdottpsi^-(\epsilon,\bdk)$ in $S_2$ and similarly
using and (\ref{gdottpsibarinter},\ref{tpsibar-intpsibark},
\ref{tpsibar+intpsibar-}) to express 
$\tpsibarbold(\epsilon,-\bdk)\cdotg$ in terms
of $\tpsibarbold^+(\epsilon,-\bdk)\cdotkh$ we obtain after some algebra
\begin{equation}
S_2 = \frac{2 a \epsilon^{-(d-2)}}{d-1+\Lambda} \int d^dk \, 
\tpsibarbold^+(\epsilon,-\bdk)\cdotkh \,
i \bdkhdotg \biggl(2k \epsilon + \frac{d(d-1-\Lambda)}{d-1} 
\frac{K_{(\Lambda-1)/2}(k \epsilon)}
{K_{(\Lambda+1)/2}(k \epsilon)}
\biggr)
\khdottpsi^-(\epsilon,\bdk)
\end{equation}
where we have dropped higher order terms in $\epsilon$ and
$K_{(\Lambda-1)/2}(k \epsilon)/
K_{(\Lambda+1)/2}(k \epsilon)$ as they will either contribute
only to contact terms or are simply higher order in $\epsilon$
than what we obtain below.  Consequently the two-point function
we obtain below is independent of the coefficient $b$. 
Substituting the expansion (\ref{ratioexp}) we again see that there
are non-trivial contributions to $S_2$ at order $\epsilon$
and at order $\epsilon^{\Lambda}$.  One may easily show
that the order $\epsilon$ terms in $S_1$ and $S_2$ cancel
so that the leading order contribution to $S'$ comes at
order $\epsilon^{\Lambda}$.  Collecting these terms and
inverse Fourier transforming we find
\begin{equation}
S' = -2 a \, c \int d^dy_1 \, d^dy_2 \bar{\psi}^{+}_i(\epsilon,\bdyo)
\frac{\gdoty1y2}{|\bdyo - \bdyt|^{d+\Lambda+1}} 
\biggl( \delta^{ij} - 2\frac{(y_1-
y_2)^i (y_1-y_2)^j}{|\bdyo-\bdyt|^2}\biggr)\psi^{-}_j(\epsilon,\bdyt)
\label{S'evaluated}
\end{equation}
where we have absorbed a factor of $\epsilon^{(\Lambda-d)/2+1}$ in both
$\psi^{-}_j(\epsilon,\bdyt)$ and $\bar{\psi}^{+}_i(\epsilon,\bdyo)$
respectively.  We therefore obtain the two-point function
\begin{equation}
\langle \xi^{+}_i(\bdyo) \bar{\xi}^{-}_j(\bdyt) \rangle = 2 a \, c
\Pi_{i}^{k} \frac{\gdoty1y2}{|\bdyo - \bdyt|^{d+\Lambda+1}} \biggl( 
\delta_{kj}
- 2\frac{(y_1-
y_2)_k (y_1-y_2)_j}{|\bdyo-\bdyt|^2}\biggr) 
\label{twopoint}
\end{equation}
where 
\begin{equation}
\Pi_{i}^j = \delta_{i}^j - \frac{1}{d}\gamma_i \gamma^j
\label{projection}
\end{equation}
projects out the transverse components to $\gamma^i$, i.e., 
$\gamma^i \Pi_{ij}=0$.
This projection operator must be present since
the sources 
$\psi^{-}_i(\epsilon,\bdx)$ and 
$\bar{\psi}^{+}_i(\epsilon,\bdx)$ are transverse to $\gamma^i$ and therefore
the boundary conformal fields are as well.

This is the expected two-point function for a pair of Rarita-Schwinger
fields of scaling dimensions $\eta=(d + \Lambda)/2$.  This follows from the
results of \cite{OP} where it is shown that the two-point function
of fields ${\cal O}^I(\bdx)$ and ${\cal O}^J(\bdy)$ of equal
scaling dimensions $\eta$ is given by 
\begin{equation}
\langle {\cal O}^I(\bdx) {\cal O}^J(\bdy) \rangle = \frac{D^{I}_{\, K}
\bigl(I(\bdx
-\bdy) \bigr) g^{KJ}}{|\bdx - \bdy|^{2 \eta}}
\end{equation}
where $D^{I}_{\, K}(R)$ form a representation of $O(d)$ (we use upper 
case Latin letters $I,J,K$ to denote an arbitrary representation)
and the metric $g^{IK}$ satisfies
\begin{equation}
D^{I_1}_{\, I_2}(R) D^{J_1}_{\, J_2}(R) g^{I_2 J_2} = g^{I_1 J_1}.
\end{equation}
$I(x)$ is the inversion operator and is given by
\begin{equation}
I_{ij}(x) = \delta_{ij} -2 \frac{x_i x_j}{|\bdx|}
\end{equation}
in the vector representation and
\begin{equation}
D(I(x)) = - \frac{\xdotg \gamma^0}{|\bdx|}
\end{equation}
in the spinor representation.
For the Rarita-Schwinger field that we are considering, the appropriate
representation is given by the direct product of the vector and
spinor representations with the transverse to $\gamma^i$ components
projected out.  Specifically, for an $O(d)$ transformation $\Lambda^{i}_{\, j}$
the corresponding element of the representation for
the Rarita-Schwinger field is $\Pi^{i}_{\,j} D(\Lambda) \Lambda^{j}_{\, k}$.
Therefore for this representation and for $\eta = (d + \Lambda)/2$ we recover
the two-point function in (\ref{twopoint}), up to an undetermined
constant.

\section{Discussion}

We have solved the boundary value problem for the massless gravitino
on the Euclidean $AdS_{d+1}$ background and via the $AdS$/CFT correspondence
\cite{gkp,wittone} have computed the two-point function of supersymmetry
currents in the dual boundary CFT.  We have found, similarly to the spinor
case \cite{HS,MuckVis2}, that the bulk solution for the massless gravitino
is given in terms of a $(d-1)$ dimensional boundary gravitino.
That is, for $\Lambda > 0 (< 0)$, Fourier
transformability of the bulk field $\psi_i(x^0,\bdx)$ implies 
that we are only free to specify on the boundary the components of 
$\psi_{i}^{-(+)}(\epsilon,\bdx)$ transverse to $\gamma^i$ while 
$\psi_{i}^{+(-)}(\epsilon,\bdx)$
necessarily vanishes in the $\epsilon \rightarrow 0$ limit.  A similar
statement holds for the adjoint gravitino with the boundary data now
consisting of the components of $\bar{\psi}_{i}^{+(-)}(\epsilon,\bdx)$
transverse to $\gamma^i$.  For $d$ odd this is
exactly as it should be since the boundary Rarita-Schwinger field, the
supersymmetry current, is transverse to $\gamma^i$
and contains half as many spinor components.
For $d$ even however the boundary supersymmetry current contains
$(d-1) \, 2^{d/2}$ components, i.e.,
effectively $(d-1)$ vector indices (the transverse to $\gamma^i$ condition
removes one vector index) but the number of spinor components
appropriate to $d$ dimensions.  The boundary data however contains
only half this number of components, and corresponds to one chirality
of the supersymmetry current.  For $d$ even therefore the two-point
function (\ref{twopoint}) is of chiral 
components of the supersymmetry current and
therefore is only half of what we want.
The other half, as noted in \cite{MuckVis2} for the spinor case, of the 
correlator for the other pair of chiral components of the supersymmetry
currents can be obtained by considering another bulk
gravitino field with $\Lambda$
having the opposite sign. 

The next natural step would be to include interactions
for the massless gravitino allowing for further checks
on the $AdS$/CFT correspondence via CFT Ward identities.
Furthermore using the Hamiltonian techniques of \cite{arufrol}
one could determine the coefficients $a$ and $b$ of the
boundary action.

\section*{Acknowledgements}
This work was supported by the Natural Sciences and Engineering
Research Council of Canada at the University of Alberta.

\end{document}